\documentclass[submission,copyright]{eptcs}
\providecommand{\event}{AFL 2014}
\usepackage{breakurl}

\usepackage{complexity}
\usepackage{graphicx} 
\usepackage{amsmath,amssymb}
\usepackage{subfigure}
\usepackage{multirow}

\usepackage{colortbl}

\newcommand{\shuffle}{\,\sqcup\!\!\!\!\!\sqcup\,}

\let\lambda=\varepsilon

\newcommand{\resp}{respectively}

\newcommand{\nfa}{\textrm{NFA}}
\newcommand{\dfa}{\textrm{DFA}}

\DeclareMathOperator{\sizec}{\textsf{size}}
\DeclareMathOperator{\rpnc}{\textsf{rpn}}
\DeclareMathOperator{\awidthc}{\textsf{awidth}}

\DeclareMathOperator{\heightc}{\textsf{height}}

\DeclareMathOperator{\dsc}{\textsf{sc}}
\DeclareMathOperator{\nsc}{\textsf{nsc}}
\DeclareMathOperator{\dtc}{\textsf{tc}}
\DeclareMathOperator{\ntc}{\textsf{ntc}}

\DeclareMathOperator{\Pos}{\textsf{Pos}}

\DeclareMathOperator{\first}{\textsf{First}}
\DeclareMathOperator{\last}{\textsf{Last}}
\DeclareMathOperator{\follow}{\textsf{Follow}}

\newcommand{\pos}{\hbox{\scriptsize\textit{pos}}}
\newcommand{\cfs}{\hbox{\scriptsize\textit{cfs}}}
\newcommand{\pd}{\hbox{\scriptsize\textit{pd}}}
\newcommand{\fol}{\hbox{\scriptsize\textit{f}}}

\newcommand*{\qed}{\raisebox{0.5ex}[0ex][0ex]{\framebox[1ex][l]{}}}
\newtheorem{theorem}{Theorem}
\newtheorem{lemma}[theorem]{Lemma}

\newtheorem{definition}[theorem]{Definition}
\newtheorem{example}[theorem]{Example}
  {\mbox{}\nolinebreak\hfill~%
  {\qed}
  \medbreak
}
\usepackage{ifpdf}
\ifpdf
\fi

\begin{document}

\title{From Finite Automata to Regular Expressions and Back---A Summary 
  on Descriptional Complexity}

\def\titlerunning{From Finite Automata to Regular Expressions and
  Back---A Summary on Descriptional Complexity}
\def\authorrunning{H.~Gruber and M.~Holzer}

\author{%
Hermann Gruber
\institute{knowledgepark AG, Leonrodstr. 68,\\
80636 M\"unchen, Germany}
\email{hermann.gruber@knowledgepark-ag.de}
\and 
Markus Holzer
\institute{Institut f\"ur Informatik, Universit\"at Giessen,\\
Arndtstr.~2, 35392 Giessen, Germany}
\email{holzer@informatik.uni-giessen.de}
}

\maketitle

\begin{abstract}
  The equivalence of finite automata and regular expressions dates
  back to the seminal paper of Kleene on events in nerve nets and
  finite automata from 1956. In the present paper we tour a fragment
  of the literature and summarize results on upper and lower bounds on
  the conversion of finite automata to regular expressions and
  \textit{vice versa}. We also briefly recall the known bounds for the
  removal of spontaneous transitions ($\lambda$-transitions) on
  non-$\lambda$-free nondeterministic devices. Moreover, we report on
  recent results on the average case descriptional complexity bounds
  for the conversion of regular expressions to finite automata and
  brand new developments on the state elimination algorithm that
  converts finite automata to regular expressions.
\end{abstract}

\section{Introduction}
\label{sec:intro}

There is a vast literature documenting the importance of the notion of
finite automata and regular expressions as an enormously valuable
concept in theoretical computer science and applications. It is well
known that these two formalisms are equivalent, and in almost all
monographs on automata and formal languages one finds appropriate
constructions for the conversion of finite automata to equivalent
regular expressions and back. Regular expressions, introduced by
Kleene~\cite{Kl56}, are well suited for human users and therefore are
often used as interfaces to specify certain patterns or languages.
For example, in the widely available programming environment
\textsc{Unix}, regular(-like) expressions can be found in legion of
software tools like, e.g., \texttt{awk}, \texttt{ed}, \texttt{emacs},
\texttt{egrep}, \texttt{lex}, \texttt{sed}, \texttt{vi}, etc., to
mention a few of them. On the other hand, automata~\cite{RaSc59}
immediately translate to efficient data structures, and are very well
suited for programming tasks. This naturally raises the interest in
conversions among these two different notions.
Our tour on the subject covers some (recent) results in the fields of
descriptional and computational complexity. During the last decade
descriptional aspects on finite automata and regular expressions
formed an extremely vivid area of research. For recent surveys on
descriptional complexity issues of finite automata and regular
expressions we refer to, for
example,~\cite{GKKLMW02,HoKu09a,HoKu10a,HoKu10,HoKu10b,HoKu11,Yu01}.
This was not only triggered by appropriate conferences and workshops
on that subject, but also by the availability of mathematical tools
and the influence of empirical studies.  For obvious reasons, this
survey lacks completeness, as finite automata and regular expressions
fall short of exhausting the large number of related problems
considered in the literature. We give a view of what constitutes, in
our opinion, the most interesting recent links to the problem area
under consideration.

Before we start our tour some definitions are in order. First of all,
our nomenclature of finite automata is as follows: a
\emph{nondeterministic finite automaton with $\lambda$-transitions}
($\lambda$-\nfa) is a quintuple $A=(Q,\Sigma,\delta,q_0,F)$, where~$Q$
is the finite set of \textit{states}, $\Sigma$ is the finite set of
\textit{input symbols}, $q_0\in Q$ is the \textit{initial state},
$F\subseteq Q$ is the set of \textit{accepting states}, and $\delta:
Q\times(\Sigma\cup\{\lambda\})\rightarrow 2^Q$ is the
\textit{transition function}. If a finite automaton has no
$\lambda$-transitions, i.e., the transition function is restricted to
$\delta:Q\times\Sigma\rightarrow 2^Q$, then we simply speak of a
\emph{nondeterministic finite automaton} (\nfa). Moreover, a
nondeterministic finite automaton is \emph{deterministic} ($\dfa$) if
and only if $|\delta(q,a)|=1$, for all states $q\in Q$ and letters
$a\in\Sigma$. 
The \emph{language accepted} by the finite automaton~$A$
is defined as $L(A) =\{\,w\in \Sigma^*\mid\mbox{$\delta(q_0,w)\cap
  F\neq\emptyset$}\,\}$, where the transition function is recursively
extended to $\delta:Q\times\Sigma^*\rightarrow 2^Q$.
Second, we turn to the definition of regular expressions: the
\emph{regular expressions} over an alphabet~$\Sigma$ and the languages
they describe are defined inductively in the usual way:\footnote{%
  For convenience, parentheses in regular expressions are sometimes
  omitted and the concatenation is simply written as
  juxtaposition. The priority of operators is specified in the usual
  fashion: concatenation is performed before union, and star before
  both product and union.} $\emptyset$, $\lambda$, and every
letter~$a$ with $a\in \Sigma$ is a regular expression, and when~$s$
and~$t$ are regular expressions, then $(s+t)$, $(s\cdot t)$, and
$(s)^*$ are also regular expressions.  The language defined by a
regular expression~$r$, denoted by $L(r)$, is defined as follows:
$L(\emptyset)=\emptyset$, $L(\lambda)=\{\lambda\}$, $L(a)=\{a\}$,
$L(s+t)=L(s)\cup L(t)$, $L(s\cdot t)=L(s)\cdot L(t)$, and
$L(s^*)=L(s)^*$. For further details on finite automata and regular
expressions we refer to, e.g.,~\cite{HoUl79}.

We start our tour on the subject with the question on the appropriate
measure for finite automata and regular expressions. We discuss this
topic in detail in Section~\ref{sec:measures}. There we also
concentrate on two specific measures: on star height for regular
expressions and cycle rank for the automaton side. By Eggan's
theorem~\cite{Eg63} both measures are related to each other. Recent
developments, in particular on the conversion from finite automata to
regular expressions, utilize this connection to prove upper and lower
bounds.
Then in Section~\ref{sec:REGEXP-to-FA} we take a closer on the
conversion from regular expressions to equivalent finite automata. We
recall the most prominent conversion algorithms such as Thompson's
construction and its optimized version the follow automaton, the
position or Glushkov automaton, and conversion by computations of the
(partial-)derivatives. We summarize the known relations on these
devices, which were mostly found during the last decade. Significant
differences on these constructions are pointed out and the presented
developments on lower bound and upper bound results enlighten the
efficiency of these algorithms. Some of the bounds are sensitive to
the size of the alphabet. Besides worst case descriptional complexity
results on the synthesis problem of finite automata from regular
expressions, we also list some recent results on the average case
complexity of the transformation of regular expressions to finite
automata.
Finally, in Section~\ref{sec:FA-to-REGEXP} we consider the converse
transformation. Again, we summarize some of the few conversion
techniques, but then stick in more detail to the so-called state
elimination technique. The reason for that is, that in~\cite{Sa09}, it
was shown that almost all conversion methods can be recast as variants
of the state elimination technique. Here, the ordering in which the
states are eliminated can largely affect the size of the regular
expression corresponding to the given finite automaton.  We survey
some heuristics that have been proposed for this goal.  For
appropriate choices of the ordering, nontrivial upper bounds on
regular expression size can be proved. By looking at the transition
structure of the \nfa{}, results from graph theory can help in
obtaining shorter expressions. There we try to illustrate the key
insights with the aid of examples, thereby avoiding the need for a
deeper dive into graph theoretic concepts. We also explain the
technique by which the recent lower bounds on regular expression size
were obtained. In this part, the known upper and lower bounds match
only in the sense that we can identify the rough order of magnitude.
So we observe an interesting tension between algorithms with provable
performance guarantees, other heuristics that are observed to behave
better in experiments, and finally some lower bounds, which seize the
expectations that we may have on practical algorithms.

\section{Measures on Finite Automata and Regular Expressions}
\label{sec:measures}

What can be said about the proper measure on finite automata and
regular expressions? For finite automata there are two commonly
accepted measures, namely the number of states and the number of
transitions.  The measure $\dsc$ ($\nsc$, \resp) counts the number of
states of a deterministic (nondeterministic, \resp) finite automaton
and $\dtc$ ($\ntc$, \resp) does the same for the number of transitions
for the appropriate devices. Moreover, $\nsc_\lambda$ ($\ntc_\lambda$,
\resp) gives the number of states (transitions, \resp) in an
$\lambda$-\nfa. The following relations between these measures are
well known---see also~\cite{MeFi71,Mo71,RaSc59}.

\begin{theorem}
 Let~$L\subseteq\Sigma^*$ be a regular language. Then 
 \begin{enumerate}
 \item $\nsc_\lambda(L)=\nsc(L)\leq\dsc(L)\leq 2^{\nsc(L)}$ and
   $\dtc(L)=|\Sigma|\cdot\dsc(L)$ and
 \item $\nsc(L)-1\leq \ntc_\lambda(L)\leq\ntc(L)\leq|\Sigma|\cdot
   (\nsc(L))^2$,
 \end{enumerate}
 where $\dsc(L)$, $\dtc(L)$ ($\nsc(L)$, $\ntc(L)$, \resp) refers to
 the minimum $\sc$ ($\nsc$, $\ntc$, \resp) among all \dfa s (\nfa s,
 \resp) accepting~$L$. Similarly, $\nsc_\lambda(L)$ ($\ntc_\lambda(L)$,
 \resp) is the minimum $\nsc_\lambda$ ($\ntc_\lambda$, \resp) among
 all $\lambda$-\nfa s for the language~$L$.
\end{theorem}

As it is defined above, deterministic transition complexity is not an
interesting measure by itself, because it is directly related
to~$\dsc$, the deterministic state complexity. But the picture changes
when deterministic transition complexity is defined in terms of
partial \dfa{}s. Here, a \emph{partial} \dfa\ is an \nfa\ which
transition function~$\delta$ satisfies $|\delta(q,a)| \le 1$, for all
states $q\in Q$ and all alphabet symbols~\mbox{$a\in \Sigma$}. A partial
\dfa\ cannot save more than one state compared to an ordinary \dfa,
but it can save a considerable number of transitions in some
cases. This phenomenon is studied, e.g., in~\cite{GSY11,MMR13,MMR13a}.
Further measures for the complexity of finite automata, in particular
measures related to unambiguity and limited nondeterminism, can be
found
in~\cite{GKKLMW02,GKW90,GLW92,HoKu10b,HKKSS02,Le98a,Le98,Le05,PSA13,PSA12,RaIb89}.

Now let us come to measures on regular expressions.  While there are
the two commonly accepted measures for finite automata, there is no
general agreement in the literature about the proper measure for
regular expressions.
We summarize some important ones: the measure $\sizec$ is defined to
be the total number of symbols (including $\emptyset$,\ $\lambda$,
symbols from alphabet~$\Sigma$, all operation symbols, and
parentheses) of a completely bracketed regular expression (for
example, used in~\cite{AHU74}, where it is called length).  Another
measure related to the reverse polish notation of a regular expression
is $\rpnc$, which gives the number of nodes in the syntax tree of the
expressions (parentheses are not counted). This measure is equal to
the length of a (parenthesis-free) expression in post-fix
notation~\cite{AHU74}.  The alphabetic width $\awidthc$ is the total
number of alphabetic symbols from~$\Sigma$ (counted with
multiplicity)~\cite{EhZe76,McNYa60}. Relations between these measures
have been studied, e.g., in~\cite{EhZe76,EKSW04,GrGu10,IlYu03}.

\begin{theorem}
 Let $L\subseteq \Sigma^*$ be a regular language. Then 
 \begin{enumerate}
 \item $\sizec(L)\leq 3\cdot \rpnc(L)$ and $\sizec(L)\leq
   8\cdot\awidthc(L)-3$,
 \item $\awidthc(L)\leq\frac12\cdot(\sizec(L)+1)$ and
   $\awidthc(L)\leq\frac12\cdot(\rpnc(L)+1)$, and 
 \item $\rpnc(L)\leq\frac12\cdot(\sizec(L)+1)$ and $\rpnc(L)\leq
   4\cdot\awidthc(L)-1$,
 \end{enumerate}
 where $\sizec(L)$ ($\rpnc(L)$, $\awidthc(L)$, \resp) refers to the
 minimum $\sizec$ ($\rpnc$, $\awidthc$, \resp) among all
 regular expressions denoting~$L$.
\end{theorem}

Further measures for the complexity of regular expressions can be
found in~\cite{BiTh10,EhZe76,EKSW04,GrJo08}. To our
knowledge, these latter measures received far less attention to date.

In the remainder of this section we concentrate on two important
measures on regular expression and finite automata that at first
glance do not seem to be related to each other: \emph{star height} and
\emph{cycle rank} or \emph{loop complexity}. Both measures are very
important, in particular, for the conversion of finite automata to
regular expressions and for proving lower bound results on the latter.
Intuitively, the star height of an expression measures the nesting
depth of Kleene-star operations. More precisely, for a regular
expression, the \emph{star height} is inductively defined by
\begin{align*}
\heightc(\emptyset) &=\heightc(\lambda)=\heightc(a)=0,\\
\heightc(s+t) &=\heightc(s\cdot t)=\max\left(\heightc(s),\heightc(t)\right),\\
\noalign{\hbox{and}}
\heightc(s^*) &= 1+\heightc(s). 
\end{align*}
The star height of a regular 
language $L$, denoted by $\heightc(L)$ 
is then defined as the minimum star height among 
all regular expressions describing~$L$. 
The seminal work dealing with the star height of regular
expressions~\cite{Eg63} established a relation between the theory of
regular languages and the theory of digraphs.  The \emph{cycle rank},
or \emph{loop complexity}, of a digraph~$D$ is defined inductively by
the following rules:~(i) the cycle rank of an acyclic digraph is zero,
(ii) cycle rank of a strongly connected component (SCC) of the digraph
with at least one arc is~$1$ plus the minimum cycle rank among the
digraphs obtainable from~$D$ by deleting a vertex, and (iii) the cycle
rank of a digraph with multiple SCCs equals the maximum cycle rank
among the sub-digraphs induced by these components.
So, roughly speaking, the cycle rank of a digraph is large if the 
cycle structure of the digraph is intricate and highly connected.
The following relation between cycle rank of automata and star height
of regular languages became known as \emph{Eggan's Theorem}~\cite{Eg63,Sa09}:

\begin{theorem}\label{thm:Eggan}
  The star height of a regular language~$L$ equals the minimum cycle
  rank among all $\lambda$-\nfa s accepting~$L$.
\end{theorem}

An apparent difficulty with applying Eggan's Theorem is that the
minimum is taken over infinitely many automata, and the cycle rank of
the minimum \dfa\ for the language does not always attain that
minimum. That makes the star height a very intricate property of
regular languages.  Indeed, the decidability status of the star height
problem was open for more than two decades, until a very difficult
algorithm was given in~\cite{Ha88a}. For recent progress on algorithms
for the star height problem, the reader is referred to~\cite{Ki05}.
From the above it is immediate that $\heightc(L) \le \nsc(L)$.  If the
language is given as a regular expression, a result from~\cite{GrHo08}
tells us a much sweeter truth:

\begin{lemma}\label{lem:star-height-lemma}
  Let $L\subseteq\Sigma^*$ be a regular language with 
  alphabetic width $n$. Then $\heightc(L) \le 3\log(n+1)$.
\end{lemma} 

The idea behind the proof of this lemma is 
that we can convert a regular expression into a 
$\lambda$-\nfa\ of similar size. The cycle structure of that 
automaton is well-behaved; and thus its cycle rank is  
low compared to the size of the automaton. Then 
Eggan's Theorem is used to convert the automaton back 
into a regular expression of low star height.

We return to the relationship between required size and 
star height of regular expressions later on. Now let us turn our 
attention to the conversion of regular expressions into equivalent
finite automata. 

\section{From Regular Expressions to Finite Automata}
\label{sec:REGEXP-to-FA}

The conversion of regular expressions into small finite automata has
been intensively studied for more than half a century.  Basically the
algorithms can be classified according to whether the output is an
$\lambda$-\nfa, \nfa, or even a \dfa. In principle one can distinguish
between the following three major construction schemes and variants
thereof:
\begin{enumerate}
\item {\em Thompson's construction}~\cite{Th68} and optimized versions, such
  as the {\em follow automaton}~\cite{IlYu03,OtFe61},
\item construction of the {\em position automaton}, or
  {\em Glushkov automaton}~\cite{Gl61,McNYa60}, and
\item computation of the {\em (partial) derivative
  automaton}~\cite{An96,Br64}.
\end{enumerate}
Further automata constructions from regular expressions can be found
in, e.g.,~\cite{ACT10,Br93,ChPa92,HSW01,PLRA11,Wa95}. We briefly
explain some of these approaches in the course of action---for further
readings on the subject we refer to~\cite{Sa09}.

\begin{figure}[tbh]
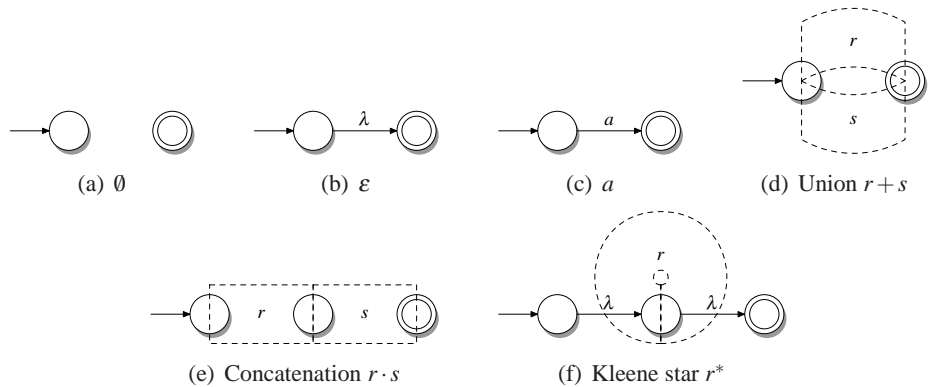

  \centering
  \subfigure[$\emptyset$]{\includegraphics[scale=0.65]{fig/fa.1}}\qquad
  \subfigure[$\lambda$]{\includegraphics[scale=0.65]{fig/fa.2}}\qquad 
  \subfigure[$a$]{\includegraphics[scale=0.65]{fig/fa.3}}\qquad 
  \subfigure[Union $r+s$]{\includegraphics[scale=0.65]{fig/fa.40}}\qquad 
  \subfigure[Concatenation $r\cdot s$]{\includegraphics[scale=0.65]{fig/fa.31}}\qquad 
  \subfigure[Kleene star $r^*$]{\includegraphics[scale=0.65]{fig/fa.32}} 
  \caption{The inductive construction of Ott and
    Feinstein~\cite{OtFe61} yielding the precursor of the follow
    automaton~$A_{\fol}(r)$ for a regular expression~$r$.}
  \label{fig:follow}
\end{figure}

Thompson's construction~\cite{Th68} was popularized by the
implementation of the \textsc{Unix} command \texttt{grep} (globally
search a regular expression and print). It amounts to the recursive
connection of sub-automata \textit{via} $\lambda$-transitions.  These
sub-automata are connected in parallel for the union, in series for the
concatenation, and in an iterative fashion for the Kleene star. This
yields an $\lambda$-\nfa{} with a linear number of states and
transitions. A structural characterization of the Thompson automaton
in terms of the underlying digraph is given in~\cite{GPWZ04,GPW99}.
Thompson's classical construction went through several stages of
adaption and optimization. The construction with the least usage of
$\lambda$-transitions was essentially given already in 1961 by Ott and
Feinstein~\cite{OtFe61}, which also can be found
in~\cite{DuKo01,Ma74a,McN82}---see Figure~\ref{fig:follow}.
Later this construction was refined by Ilie and Yu~\cite{IlYu03}
and promoted under the name \emph{follow automaton}.
In fact, the follow automaton is constructed from a regular
expression~$r$ by recursively applying the construction of Ott and
Feinstein and simultaneously improving on the use of
$\lambda$-transitions in the following sense: (i) in the concatenation
construction a $\lambda$-transition into the common state to both
sub-automata leads to an appropriate state merging; similarly a state
merging is done for an $\lambda$-transition leaving the common state,
(ii) in the Kleene star construction, if the middle state is on a
cycle of $\lambda$-transitions, all these transitions are removed, and
all states of the cycle are merged, and (iii) after the construction is
finished, a possible $\lambda$-transition from the start state is
removed and both involved states are merged appropriately. Notice,
that the automaton thus constructed may still contain
$\lambda$-transitions. In order to amend the situation, an
$\lambda$-removal procedure is applied: simply replace any sequence of
an $\lambda$-transition followed by an $a$-transition by directly
connecting the states on both ends of the sequence by a single
$a$-transition directly. A final step takes care about the
$\lambda$-transition to the final state. This results in the follow
automaton~$A_{\fol}(r)$ of~\cite{IlYu03}, for the regular expression~$r$.

\begin{example}\label{exa:software-buffer}
  Imagine a software buffer supporting the actions~$a$ (``add work
  packet'') and~$b$ (``remove work packet''), with a total capacity
  of~$n$ packets. Let~$r_n$ denote the regular expression for the
  action sequences that result in an empty buffer and never cause the
  buffer to exceed its capacity. Then
  $$r_1=(ab)^*\quad\mbox{and}\quad r_n=(a\cdot r_{n-1}\cdot b)^*,\quad\mbox{for $n\geq 2$.}$$
  Following the construction of the follow automaton as described
  in~\cite{IlYu03} results in the automaton depicted in
  Figure~\ref{fig:follow-for-Rn}.
 \begin{figure}[htb]
   \centering
   \includegraphics[scale=0.65]{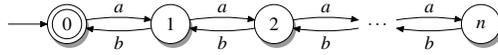}
   \caption{The follow automaton~$A_{\fol}(r_n)$ accepting $L(r_n)$.}
   \label{fig:follow-for-Rn}
 \end{figure}
 Observe the constructed automaton is minimal, which is not the case
 in general. This is our running example, where the behaviour of the
 state elimination technique described in the next section is
 discussed in more detail.\hfill\qed
\end{example}

Preliminary bounds on the required size of a finite automaton
equivalent to a given regular expression were given in~\cite{IlYu03}.
Later, a tight bound in terms of reverse polish
notation~\cite{GuFe08}, and also a tight bound in terms of alphabetic
width was found~\cite{GrGu10}.  In the next theorem we summarize the
results from~\cite{GrGu10,GuFe08,IlYu03}---here size of an automaton
refers to the sum of the number of states and the number transitions:

\begin{theorem}
  Let $n\ge 1$, and~$r$ be a regular expression of alphabetic
  width~$n$. Then size~$\frac{22}{5}n$ is sufficient for an equivalent
  $\lambda$-\nfa\ accepting~$L(r)$. In terms of reverse polish
  length, the bound is $\frac{22}{15}(\rpnc(r)+1)+1$.  Furthermore,
  there are infinitely many languages for which both bounds are tight.
\end{theorem}

The aid for the tight bound in terms of the alphabetic width stated in
the previous theorem is a certain normal form for regular expressions,
which is a refinement of the \emph{star normal form} from~\cite{Br93}.
The definition reads as follows---transformation into strong star normal form
preserves the described language, and is weakly monotone with respect
to all usual size measures:

\begin{definition}\label{defn:senf}
  The operators $\circ$ and $\bullet$ are defined on regular
  expressions\footnote{%
    Since~$\emptyset$ is only needed to denote the empty set, and the
    need for~$\lambda$ can be substituted by the operator $L^? = L
    \cup \{\lambda\}$, an alternative is to introduce also the
    $^?$-operator and instead forbid the use of~$\emptyset$
    and~$\lambda$ inside non-atomic expressions.  This is sometimes
    more convenient, since one avoids unnecessary redundancy already
    at the syntactic level~\cite{GrGu10}.}  over
  alphabet~$\Sigma$. The first operator is given by: $a^\circ = a$,
  for $a\in \Sigma$,\ $(r+s)^\circ = r^\circ + s^\circ$,\ $r^{?\circ}
  = r^\circ$,\ $r^{*\circ} =r^\circ$; finally, $(r\cdot s)^\circ=
  r\cdot s$, if $\lambda\notin L(rs)$ and $r^\circ + s^\circ$
  otherwise.  The second operator is given by: $a^\bullet = a$, for
  $a\in \Sigma$,\ $(r+s)^\bullet = r^\bullet + s^\bullet$,\ $(r\cdot
  s)^\bullet = r^\bullet\cdot s^\bullet$,\ $r^{*\bullet} =
  r^{\bullet\circ*}$; finally, $r^{?\bullet} = r^{\bullet}$, if
  $\lambda\in L(r)$ and $r^{?\bullet} = r^{\bullet?}$ otherwise.  The
  \emph{strong star normal form} of an expression~$r$ is then defined
  as~$r^\bullet$.
\end{definition}

What about the transformation of a regular expression into a finite
automaton if $\lambda$-transitions are not allowed?  One way to obtain
an \nfa\ directly is to perform the standard algorithm for removing
$\lambda$-transitions, see, e.g.,~\cite{HoUl79}, which may increase
the number of transitions at most quadratically. Another way is to
directly implement the procedure during the recursive construction
using non-$\lambda$-transitions to connect the sub-automata
appropriately. Constructions of this kind can be found in,
e.g.,~\cite{AhUl72,Le81a}. For the conversion of $\lambda$-\nfa s to
\nfa s the lower bound of~\cite{HrSc05} applies. There it was shown
that there are infinitely many languages which are accepted by
$\lambda$-\nfa s with $O(n\cdot(\log n)^2)$ transitions, such that any
\nfa\ needs at least $\Omega(n^2)$ transitions. This lower bound is
witnessed by a language over a growing size alphabet and shows that,
in this case, the standard algorithm for removing
$\lambda$-transitions cannot be improved significantly. For the case
of binary alphabets, a lower bound of $\Omega(n\cdot
2^{c\cdot\sqrt{\log n}})$, for every $c<\frac12$, was proved
in~\cite{HrSc05} as well.

Another possibility to obtain ordinary \nfa s is to directly construct
the \emph{position automaton}, also called the \emph{Glushkov
  automaton}~\cite{Gl61}---see also~\cite{McNYa60}. Intuitively, the
states of this automaton correspond to the alphabetic symbols or, in
other words, to positions between subsequent alphabetic symbols in the
regular expression. Let us be more precise: assume that~$r$ is a
regular expression over~$\Sigma$ of alphabetic width~$n$. In~$r$ we
attach subscripts to each letter referring to its position (counted
from left to right) in~$r$. This yields a \emph{marked}
expression~$\overline{r}$ with distinct input symbols over an alphabet
$\overline{\Sigma}$ that contains all letters that occur
in~$\overline{r}$.  To simplify our presentation we assume that the
same notation is used for unmarking, i.e.,
$\overline{\overline{r}}=r$.  Then in order to describe the position
automaton we need to define the following sets of positions on the
marked expression. Let $\Pos(r)=\{1,2,\ldots,\awidthc(r)\,\}$ and
$\Pos_0(r)=\Pos(r)\cup\{0\}$.  The position set $\first$ takes care of
the possible beginnings of words in~$L(\overline{r})$. It is
inductively defined as follows:
\begin{align*}
 \first(\emptyset) & = \first(\lambda)=\emptyset,\\
 \first(a_i) &=\{i\},\\ 
 \first(s+t) &=\first(s)\cup\first(t),\\
 \first(s\cdot t) &=\begin{cases} 
   \first(s)\cup\first(t) & \mbox{if $\lambda\in L(s)$}\\
   \first(s) & \mbox{otherwise,}
\end{cases}\\
\noalign{\hbox{and}}
 \first(s^*) &=\first(s).
\end{align*}
Accordingly the position set $\last$ takes care of the possible
endings of words in~$L(\overline{r})$. Its definition is similar to
the definition of $\first$, except for the concatenation, which reads
as follows: 
\begin{align*}
\last(s\cdot t) &=
\begin{cases}
 \last(s)\cup\last(t) & \mbox{if $\lambda\in L(t)$}\\
 \last(t) & \mbox{otherwise.} 
\end{cases} 
\end{align*}
Finally, the set $\follow$ takes care about the possible continuations
in the words in~$L(\overline{r})$. It is inductively defined as
\begin{align*}
 \follow(\emptyset) & =\follow(\lambda)=\follow(a_i)=\emptyset\\ 
 \follow(s+t) & =\follow(s)\cup\follow(t)\\
 \follow(s\cdot t) & =\follow(s)\cup\follow(t)\cup\last(s)\times\first(t)\\
\noalign{\hbox{and}}
 \follow(s^*) & =\follow(s)\cup\last(s)\times\first(s).
\end{align*}
Then the position automaton for~$r$ is defined as
$A_{\pos}(r)=(\Pos_0(r),\Sigma,\delta_{\pos},0,F_{\pos})$, where
$\delta(0,a)=\{\,j\in\first(\overline{r})\mid a=\overline{a_j}\,\}$,
for every $a\in\Sigma$ and
$\delta(i,a)=\{\,j\mid\mbox{$(i,j)\in\follow(\overline{r})$ and
  $a=\overline{a_j}$}\,\}$, for every $i\in\Pos(r)$ and $a\in\Sigma$,
and $F_{\pos}=\last(\overline{r})$, if $\lambda\not\in L(r)$, and
$F_{\pos}=\last(\overline{r})\cup\{0\}$ otherwise.

\begin{example}
  Consider the regular expression~$r_n$ from
  Example~\ref{exa:software-buffer}. If we mark the regular
  expression~$r_n$, then we obtain $\overline{r_n}=(a_1(a_2(a_3\ldots
  b_{2n-2})^*b_{2n-1})^*b_{2n})^*$.
 Easy calculations show that the position sets read as follows:
  \begin{align*}
    \first(\overline{r_n}) & = \{1\}\\
    \last(\overline{r_n}) & = \{2n\}\\
    \noalign{\hbox{and}} 
   \follow(\overline{r_n}) & = \{\,(i,i+1)\mid
    1\leq i < 2n\,\}\cup\{\,(i,2n-i+1),(2n-i+1,i)\mid 1\leq i\leq
    n\,\}
  \end{align*}
  The position automaton on state set $\Pos_0(r)$ is depicted in
  Figure~\ref{fig:pos-for-Rn}. Here the set of final states is
  $F_{\pos}=\{0,2n\}$, since $\lambda\in L(r_n)$.
  \begin{figure}[htb]
    \centering
    \includegraphics[scale=0.65]{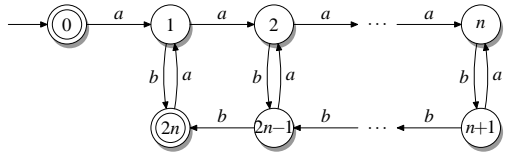}
    \caption{The position automaton~$A_{\pos}(r_n)$ accepting $L(r_n)$.}
    \label{fig:pos-for-Rn}
  \end{figure}
  Observe, that the follow automaton~$A_{\fol}(r_n)$ can be obtained
  from~$A_{\pos}(r_n)$ by taking the quotient of automata, i.e.,
  merging of states, with respect to the relation $\equiv_f$ described
  in~\cite{IlYu03}, which contains the elements $(i,2n-i)$, for $0\leq
  i\leq 2n$. This leads to the merging of states~$0$ and~$2n$,
  states~$1$ and~$2n-1$, states~$2$ and~$2n-2$, up to states~$n-1$
  and~$n+1$.\hfill\qed
\end{example}

An immediate advantage of the position automaton is observed, e.g.,
in~\cite{ASU86,BeSe86}: for a regular expression~$r$ of alphabetic
width~$n$, for $n\ge 0$, the position automaton~$A_{\pos}(r)$ always
has precisely~$n+1$ states. Simple examples, such as the singleton set
$\{a^n\}$, show that this bound is tight. Nevertheless, several
optimizations have been developed that give \nfa s having often a
smaller number of states, while the underlying constructions are
mathematically sound refinements of the basic construction. A
characterization of the position automaton is given
in~\cite{CaZi00}. Moreover, structural comparisons between the
position automaton with its refined versions, namely the \emph{follow
  automaton}, the \emph{partial derivative automaton}~\cite{An96}, or
the \emph{continuation automaton}~\cite{BeSe86} is given
in~\cite{ChZi02,IlYu03}. The partial derivative automaton is known
under different names, such as \emph{equation automaton}~\cite{Mi66}
or \emph{Antimirov automaton}~\cite{An96}. Further results on
structural properties of these automata, when built from regular
expressions in star normal form, can be found in~\cite{COZ07,Ch10}.
A quantitative comparison on the sizes of the the aforementioned \nfa
s for specific languages shows that they can differ a lot. The results
listed in Table~\ref{tab:comparison} are taken
from~\cite{IlYu03}---here size of an automaton refers to the sum of
the number of states and the number transitions.
\def\shade{\cellcolor[gray]{0.75}}%
\begin{table}[bth]
\renewcommand{\arraystretch}{1.5}%
  \centering
  \begin{tabular}{|l||c|c|c|c|}\hline\hline
    & \multicolumn{4}{c|}{Finite Automaton}\\\cline{2-5}
Expression & $A_{f}(\cdot) $ & $A_{\pd}(\cdot) $ & $A_{\pos}(\cdot) $ & $A_{\cfs}(\cdot) $\\\hline\hline
$r_1=(a_1+\lambda)^*$ and $r_{n+1}=(r_n +s_n)^*$ & \shade & \multicolumn{2}{c|}{\multirow{2}{*}{$\Theta(|r_n|^2)$}} & \multirow{2}{*}{$\Theta(|r_n|\cdot(\log |r_n|)^2)$} \\
with $s_n=r_n[a_j\mapsto a_{j+2^{n-1}}]$ & \multirow{-2}{*}{\shade $\Theta(|r_n|)$} & \multicolumn{2}{c|}{} & \\\hline 
$r_{n,m}=(\sum_{i=1} a_i)(\sum_{i=1}^n a_i+\sum_{i=1}^m b_i)^*$ & \multicolumn{2}{c|}{\shade $\Theta(|r_{n,m}|)$} & $\Theta(|r_{n,m}|^2)$ & $\Theta(|r_{n,m}|\cdot(\log |r_{n,m}|)^2)$\\\hline
$r_n=\sum_{i=1}^na_i\cdot (b_1+b_2+\ldots+b_n)^*$ & $\Theta(|r_n|)$ & \shade $\Theta(|r_n|^{1/2})$ & $\Theta(|r_n|^{3/2})$ & $\Theta(|r_n|\cdot(\log |r_n|)^2)$\\\hline
$r_n=(a_1+\lambda)\cdot (a_2+\lambda)\cdots(a_n+\lambda)$ & \multicolumn{3}{c|}{$\Theta(|r_n|^2)$} & \shade $\Theta(|r_n|\cdot(\log |r_n|)^2)$\\\hline 
  \end{tabular}
  \caption{Comparing sizes of some automata constructions for specific languages from the literature---gray shading marks the smallest automaton. Here~$A_{\fol}$ refers to the follow automaton, $A_{\pd}$ to the partial derivative automaton, $A_{\pos}$ to the position automaton, and $A_{\cfs}$ to the common follow set automaton. Moreover,~$|r_n|$ ($|r_{n,m}|$, \resp) refers to the alphabetic width of the regular expression~$r_n$ ($r_{n,m}$, \resp).}
  \label{tab:comparison}
\end{table}
For comparison reasons also the \emph{common follow set
  automaton}~$A_{\cfs}$ is listed---since the description of~$A_{\cfs}$
is quite involved we refer the reader to~\cite{HSW01}.  There, this
automaton was used to prove an upper bound on the number of
transitions. The issue on transitions for \nfa s, in particular when
changing from an $\lambda$-\nfa\ to an \nfa , is discussed next.

\begin{figure}[htb]
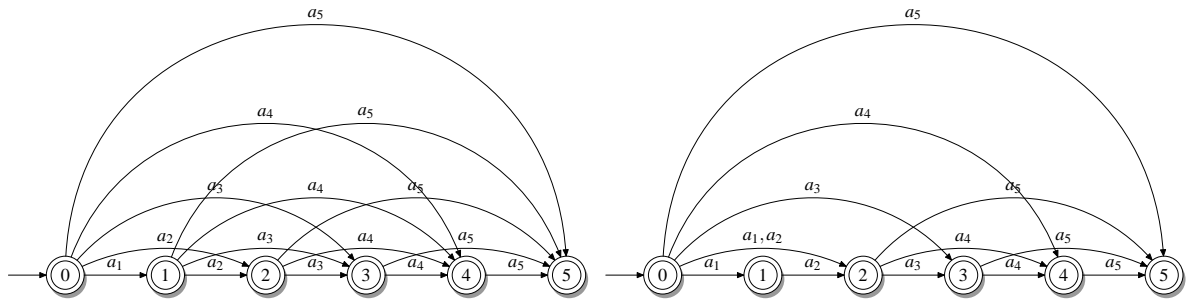

  \centering
\hspace*{\fill}%
\includegraphics[scale=0.63]{fig/fa2.100}
\hfill%
\includegraphics[scale=0.63]{fig/fa2.101}
\hspace*{\fill}%
\caption{Let $r_n=(a_1+\lambda)\cdot (a_2+\lambda)\cdots(a_n+\lambda)$
  and $n=5$. Position automaton~$A_{\pos}(r_5)$ (left) and its refined
  version the common follow set automaton~$A_{\cfs}(r_5)$ (right)
  accepting language~$L(r_5)$; in both cases the dead state and all
  transitions leading to it are not shown. The
  automaton~$A_{\cfs}(r_5)$ is obtained as follows: the state~$1$
  of~$A_{\pos}(r_5)$ is split such that the new state gets the outgoing
  transitions labeled with~$a_3$,\ $a_4$, and~$a_5$, and is finally
  identified with state~$2$, which can be done since it has the same
  outgoing transitions.}
  \label{fig:HSW01}
\end{figure}

Despite the mentioned optimizations, except for the common follow set
automaton, all of these constructions share the same problem with
respect to the number of transitions.  An easy upper bound on the
number of transitions in the position automaton is $O(n^2)$,
independent of alphabet size.  It is not hard to prove that the
position automaton for the regular expression $$r_n =
(a_1+\lambda)\cdot(a_2+\lambda)\cdots(a_n+\lambda)$$ has $\Omega(n^2)$
transitions.  It appears to be difficult to avoid such a quadratic
blow-up in actual size if we stick to the \nfa\ model. Also if we
transform the expression first into a $\lambda$-\nfa\ and perform the
standard algorithm for removing $\lambda$-transitions, see,
e.g.,~\cite{HoUl79}, we obtain no better result.  This naturally
raises the question of comparing the descriptional complexity of
$\nfa$s over regular expressions.  For about forty years, it appears
to have been considered as an unproven factoid that a quadratic number
of transitions will be inherently necessary in the worst case
(cf.~\cite{HSW01}).  A barely super-linear lower bound of $\Omega(n\log
n)$ on the number of transitions of any \nfa\ accepting the language
of the expression~$r_n$ was proved~\cite{HSW01}.  More interestingly,
the main result of that paper is an algorithm transforming a regular
expression of size~$n$ into an equivalent $\nfa$ with at most
$O(n\cdot(\log n)^2)$ transitions. See Figure~\ref{fig:HSW01} on how
the algorithm of~\cite{HSW01} saves transitions for regular
expression~$r_n$, explained for $n=5$.
In fact, this upper bound made
their lower bound look reasonable at once!  Shortly thereafter, an
efficient implementation of that conversion algorithm was
presented~\cite{HaMu00}, and the lower bound was improved in~\cite{Li03}
to~$\Omega(n\cdot(\log n)^2/\log \log n)$. Later work~\cite{Sch06}
established that any \nfa\ accepting language~$L(r_n)$ indeed must
have at least $\Omega(n\cdot (\log n)^2)$ transitions. So the upper
bound of $O(n\cdot(\log n)^2)$ from~\cite{HSW01} is asymptotically
tight:

\begin{theorem}
  Let $n\ge 1$ and~$r$ be a regular expression of alphabetic
  width~$n$. Then $O(n\cdot(\log n)^2)$ transitions are sufficient for
  an \nfa\ to accept~$L(r)$.  Furthermore, there are infinitely many
  languages for which this bound is tight.
\end{theorem}
 
Notice that the example witnessing the lower bound is over an alphabet
of growing size. For alphabets of size two, the upper bound was
improved first~\cite{Ge03} to~$O(n\cdot\log n)$, and then even
to~$n\cdot 2^{O(\log^*n)}$, where $\log^*$ denotes the iterated binary
logarithm~\cite{Sch06}. Moreover, a lower bound of $\Omega(
n\cdot(\log k)^2 )$ on the size of \nfa s with $k$-letter input
alphabet was show in~\cite{Sch06}, too. Thus the question from~\cite{Hr02} 
whether a conversion from regular expressions over a
binary alphabet into \nfa s of linear size is possible, is almost
settled by now.

\begin{theorem}
  Let $n\ge 1$ and~$r$ be a regular expression of alphabetic width~$n$
  over a binary alphabet. Then~$n\cdot 2^{O(\log^*n)}$ transitions are
  sufficient for a \nfa\ to accept~$L(r)$.
\end{theorem}

Next, let us briefly discuss the problem of converting regular
expressions to \dfa s. Again, this problem has been studied by many
authors.  The obvious way to obtain a \dfa\ is by applying the well
known \emph{subset} or \emph{power-set construction}~\cite{RaSc59}. Due
to this construction the obtained \dfa\ may be of exponential size. A
more direct and convenient way is to use Brzozowski's derivatives of
expressions~\cite{Br64}.  A taxonomy comparing many different
conversion algorithms is given in~\cite{Wa95}. Regarding the
descriptional complexity, a tight bound of $2^{n}+1$ states in terms
of alphabetic width is given in~\cite{Le81a}. The mentioned work also
establishes a matching lower bound, but for a rather nonstandard
definition of size. In terms of alphabetic width, the best lower bound
known to date is from~\cite{EKSW04}.  Together, we have the following
result:

\begin{theorem}
  Let $n\ge 1$ and~$r$ be a regular expression of alphabetic width~$n$
  over a binary alphabet. Then $2^{n}+1$ states are sufficient for a
  \dfa\ to accept~$L(r)$.  In contrast, for infinitely many $n$ there
  are regular expressions $r_n$ of alphabetic width~$n$ over a binary
  alphabet, such that the minimal \dfa\ accepting $L(r_n)$ has at
  least $\frac54 2^{\frac{n}{2}}$ states.
\end{theorem}

Recent developments on the conversion of regular expressions to finite
automata show an increasing attention on the study of descriptional
complexity in the average case. For instance, in~\cite{Ni09} it was
shown that, when choosing the expression uniformly at random, the
position automaton has $\Theta(n)$ transitions on average, where~$n$
refers to the nodes in the parse tree of the expression. A similar
result holds w.r.t.\ alphabetic width, for the position automaton as
well as for the partial derivative automaton~\cite{BMMR12}. A closer
look reveals that the number of transitions in the partial derivative
automaton is, on average, half the size of the number of transitions
in the position automaton~\cite{BMMR12}, for large alphabet sizes;
this also holds for the number of states~\cite{BMMR10}. Results on the
average size of $\lambda$-\nfa s built from Thompson's construction
and variants thereof~\cite{IlYu03,SiSo88,Th68} can be found
in~\cite{BMMR14}---in their investigation the authors consider the
follow automaton before the final $\lambda$-removal is done. Let us
call this device \emph{$\lambda$-follow automaton}. It turns out that
the $\lambda$-follow automaton is superior to the other constructions
considered. In particular, the number of $\lambda$-transitions
asymptotically tends to zero, i.e., the $\lambda$-follow automaton
approaches the follow-automaton.

Almost all of these results were obtained with the help of the
framework of analytic combinatorics~\cite{FlSe09}. The idea to use
this approach is quite natural. Recall, that the number of regular
expressions of a certain size measured by, e.g., alphabetic width, can be
counted by using generating functions---for more involved measures, one
has to use multivariate generating functions.  To this end one
transforms a grammar describing regular expressions such as, e.g., the
grammar devised in~\cite{GLS12}, into a generating function. Since
the grammar describes a combinatorial class, the generating function
can be obtained by the \emph{symbolic method} of~\cite{FlSe09}, and the
coefficients of the power series can be estimated to give
approximations of the measure under consideration. 

Finally, let us
note, that the results on the average size of automata depends on the
probability distribution that is used for the average-case analysis. 
In~\cite{NPR10} it was shown that
the number of transitions of the position automaton is
in~$\Theta(n^2)$ under a distribution that is inspired from random
binary search trees (BST-like model). To our knowledge, average case
analysis under the BST-like model for other automata such as the
follow automaton or the partial derivative automaton, has not been
conducted so far.

\section{From Finite Automata to Regular Expressions}
\label{sec:FA-to-REGEXP}

There are a few classical algorithms for converting finite automata
into equivalent regular expressions, namely
\begin{enumerate}
\item the \emph{algorithm based on Arden's lemma}~\cite{Ar61,Co71e}, and
\item the \emph{McNaughton-Yamada algorithm}~\cite{McNYa60}, and 
\item the \emph{state elimination technique}~\cite{BrMcC63}.
\end{enumerate}
These procedures look different at first
glance. We briefly explain the main idea of these 
approaches---for a detailed description along with an 
explanation of the differences between the methods, 
the reader is referred to~\cite{Sa09}.
There it is shown, that all of the above approaches are more or less
reformulations of the same underlying algorithmic idea, and they yield
(almost) the same regular expressions.\footnote{%
  Let us also mention that there is another algebraic algorithm
  from~\cite{Co71e}, which is based on the recursive decomposition of
  matrices into blocks. Here, the precise relation to the
  aforementioned algorithms remains to be investigated~\cite{Sa09}.}

An algebraic approach to solve the conversion problem from finite
automata to regular expressions is the \emph{algorithm based on
  Arden's lemma}~\cite{Ar61,Co71e}.  It puts forward a set of language
equations for a given finite automaton. Here, the $i$th equation
describes the set $X_i$ of words $w$ such that the given automaton can
go from the $i$th state to an accepting state on reading $w$. That
system of equations can be resolved by eliminating the indeterminates
$X_i$ using a method that resembles Gaussian elimination. But we work
in a an algebraic structure different from a field, so for the
elimination of variables, we have to resort to \emph{Arden's lemma}:

\begin{lemma}
  Let $\Sigma$ be an alphabet, and let $K, L \subseteq \Sigma^*$,
  where $K$ does not contain the empty word~$\lambda$.  Then the set
  $K^*L$ is the unique solution to the language equation $X = K\cdot X
  + L$, where $X$ is the indeterminate.
\end{lemma} 

Now let us have a look on how Arden's lemma can be applied to our
running example.

\begin{example}
  From the automaton depicted in Figure~\ref{fig:follow-for-Rn} one
  reads off the equations
$$
X_0 =a\cdot X_1 + \lambda, \quad X_i =a\cdot X_{i+1}+ b\cdot
X_{i-1},\quad\mbox{for $1\leq i< n$, and}\quad X_{n} =b\cdot
X_{n-1}.$$ Substituting the right hand side of $X_n$ in the next to
last equation and solving it by Arden's lemma results in
$X_{n-1}=(ab)^*b\cdot X_{n-2}$. For short, $X_{n-1}=r_1\cdot b\cdot
X_{n-2}$, where $r_i$ is defined as in
Example~\ref{exa:software-buffer}. Next this solution is substituted
into the equation for~$X_{n-2}$. Solving for~$X_{n-2}$ gives us
$X_{n-2}=r_2\cdot b\cdot X_{n-3}$. Proceeding in this way up to the
very first equation gives us $X_0=a\cdot r_{n-1}\cdot b\cdot
X_0+\lambda$. The solution to the indeterminate~$X_0$ is according to
Arden's lemma $(a\cdot r_{n-1}\cdot b)^*\cdot\lambda=r_n$, by applying
obvious simplifications.  Hence, for instance, in case $n=6$ we obtain
$(a(a(a(a(a(ab)^*b)^*b)^*b)^*b)^*b)^*$.\hfill\qed
\end{example}

The \emph{McNaughton-Yamada algorithm}~\cite{McNYa60}
maintains a matrix with regular expression entries, 
where the rows and columns are the states of the given 
automaton. The iterative algorithm uses a ranking 
on the state set, and proceeds in $n$ rounds,
if $n$ is the number of states in the given automaton~$A$.
In the matrix $(a_{jk})_{j,k}$ computed in round~$i$, 
the entry $a_{jk}$ is an expression describing the 
nonempty labels $w$ of computations of $A$ starting
in $j$ and ending in $k$, such that none of the 
intermediate states of the computation is ranked  
higher than~$i$. From these expressions, it is 
not difficult to obtain a regular expression 
describing~$L(A)$.

\begin{example}
Running the McNaughton-Yamada algorithm on the 
automaton depicted in Figure~\ref{fig:follow-for-Rn} 
for $n=3$ with the ranking $3,2,1,0$
starts with the following matrix:
\[
\bordermatrix{ 
  & 3         & 2 & 1 & 0 \cr
3 & \emptyset & b         & \emptyset & \emptyset \cr
2 & a         & \emptyset & b         & \emptyset \cr
1 & \emptyset & a         & \emptyset & b         \cr
0 & \emptyset & \emptyset & a         & \emptyset \cr
}
\]
If $(a_{jk})_{j,k}$ denotes the matrix computed in 
round $i$, then the matrix $(b_{jk})_{j,k}$ for round
$i+1$ can be computed using the rule
\[b_{jk} = a_{jk} + a_{ji}(a_{ii})^*a_{ik}\]
After the first round, the entry in the upper left corner 
of the matrix reads as $\emptyset + \emptyset \emptyset^* \emptyset$.
It is of course helpful to simplify the intermediate regular
expressions, by applying some obvious simplifications. 
As noted in~\cite{McNYa60}, we can use in particular
\[b_{ij} = (a_{ii})^*a_{ij}\quad\mbox{and}\quad b_{ji}=a_{ji}(a_{ii})^*.\]
Then the matrix computed in the first round reads as
\[
\begin{pmatrix} 
\emptyset & b         & \emptyset & \emptyset \\
a         & ab        & b         & \emptyset \\
\emptyset & a         & \emptyset & b         \\
\emptyset & \emptyset & a         & \emptyset \\ 
\end{pmatrix},
\]
the one from the second round is
\[
\begin{pmatrix}
b(ab)^*a  & b(ab)^*    & b(ab)^*b & \emptyset \\
(ab)^*a   & (ab)^*ab   & (ab)^*b  & \emptyset \\
a(ab)^*a  & a(ab)^*    & a(ab)^*b & b         \\
\emptyset & \emptyset & \emptyset & \emptyset \\
\end{pmatrix},
\]
and the computation is continued in the same vein. 
Finally, the entry in the lower-right corner of the matrix reads as $(a(a(ab)^*b)^*b)^*a(a(ab)^*b)^*b$, 
and the desired regular expression describing $L_3$ is obtained by adding the empty word: $\lambda+(a(a(ab)^*b)^*b)^*a(a(ab)^*b)^*b$.
\end{example}

A few industrious readers, who have worked out the calculation of the
previous example until the final matrix, may have observed that many
of the intermediate expressions were actually not needed for the final
result. Indeed, in a computer implementation~\cite[page~8]{McI68} of
the basic McNaughton-Yamada algorithm during the $1960$s, the author
notes: \textit{``a basic fault of the method is that it generates such
  cumbersome and so numerous expressions initially.''}  Below we 
discuss how the generation of unnecessary sub-expressions can be
avoided.

We now come to an algorithm that we describe in greater detail, namely
the \emph{state elimination algorithm}~\cite{BrMcC63}.  This procedure
maintains an extended finite automaton, whose transitions are labeled
with regular expressions, rather than alphabet symbols.
The computation of an \nfa\ $A$ can be thought of as reading the input
word letter by letter, thereby nondeterministically changing its state
with each letter in a way that is consistent with its transition
table~$\delta$. On reading a word $w\in \Sigma$, we say that the
finite automaton~$A$ and can go on input~$w$ from state~$j$ to
state~$k$, if there is a computation on input~$w$ taking~$A$ from
state~$j$ to~$k$. Similarly, for a subset~$U$ of the state set~$Q$ of
the automaton~$A$, we say that~$A$ can go on input~$w$ from state~$j$
\emph{through~$U$} to state~$k$, if there is a computation on
input~$w$ taking~$A$ from state~$j$ to~$k$, without going through any
state outside~$U$, except possibly~$j$ and~$k$.  With the {\it
  r\^oles} of~$j$, $k$, and~$U$ fixed as above, we now define the
language~$L_{jk}^U$ as the set of input words on which the
automaton~$A$ can go from~$j$ to~$k$ through~$U$.
The state elimination scheme fixes an ordering on the state set~$Q$.
Starting with $U = \emptyset$, regular expressions denoting the
languages $L_{jk}^\emptyset$ for all pairs $(j,k) \in Q\times Q$ can
be easily read off from the transition table of~$A$.  Now an important
observation is that for each state $i \in Q\setminus U$ holds
\begin{equation*}
L_{jk}^{U\cup \{i\}} = L_{jk}^U \cup  L_{ji}^U\cdot (L_{ii}^{U})^*\cdot L_{ik}^U.
\end{equation*}
Letting~$i$ run over all states according to the ordering, we can grow
the set~$U$ one by one, in each round computing the intermediate
expressions $r^{U\cup\{i\}}_{jk}$ for all~$j$ and~$k$. The final
regular expression is obtained by utilizing the fact
$L(A)=\bigcup_{f\in F} L_{q_0f}^Q$.

As observed already by McNaughton and Yamada~\cite{McNYa60}, we have
$\awidthc(L^\emptyset_{jk}) \le |\Sigma|$, and each round increases
the alphabetic width of each intermediate sub-expression by a factor of
at most~$4$.  Another convenient trick is to modify the automaton, by
adding a new initial state~$s$ and a new final state~$t$ to the
automaton without altering the language, such that $t$ is the single
final state, and there are no transitions entering $s$ or leaving~$t$.
Then $s$ and $t$ need not to be added to the set~$U$. Instead, observe
that $L(A)=L_{st}^U$, with $U = Q\setminus\{s,t\}$.  We also
note\footnote{The same trick applies for the McNaughton-Yamada
  algorithm: If the single initial and the single final state are not
  eliminated, we can erase the entries of the $i$th row and the $i$th
  column of the computed matrix in round $i$.  } that the computation
of~$r_{jk}^U$ needs to be carried out only for those~$j$ and~$k$ not
in $U$.  We thus obtain the following bound:

\begin{theorem}
 Let $n\ge 1$ and~${A}$ be an $n$-state \nfa\ over
  alphabet~$\Sigma$. Then alphabetic width $|\Sigma|\cdot 4^{n}$ 
  is sufficient for
  a regular expression describing~$L({A})$. Such an expression can be
  constructed by state elimination.
\end{theorem}

In contrast, the state elimination algorithm might suddenly yield a 
much simpler regular expression 
once we change the ordering in which the states are eliminated. 
We illustrate the influence of
the elimination ordering on a small example. 

\begin{example}\label{ex:expression-size-height}
  Consider our software buffer from Example~\ref{exa:software-buffer}
  for~$n=6$. Let $L_n:=L(r_n)$. For illustration, a minimal \dfa\
  for~$L_6$ is depicted in Figure~\ref{fig:DFA-buffer}.
The two regular expressions 
\[(a(a(a(a(a(ab)^*b)^*b)^*b)^*b)^*b)^*\]
and 
\begin{equation*}
\lambda+a(ab+ba)^*b+a(ab+ba)^*aa\left(ab+ba+bb(ab+ba)^*aa+aa(ab+ba)^*bb\right)^*bb(ab+ba)^*b
\end{equation*}
both describe the language~$L_6$. 
The first expression is obtained by eliminating the states in the
order $6$, $5$, $4$, $3$, $2$, $1$, and $0$, while the second
expression is produced by the order $0$, $2$, $4$, $6$, $1$, $5$,
and~$3$.
\begin{figure}[htb]
  \centering 
  \hfill\includegraphics[scale=0.65]{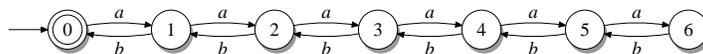}\hspace*{\fill}
  \caption{A minimal \dfa\ accepting the language~$L_6:=L(r_6)$.}
  \label{fig:DFA-buffer}
\end{figure}
Note that the expressions have very different structure. 
The first is much shorter, but has star height~$6$,
while the second, and longer expression, has star height~$2$. 
Indeed, in~\cite{McN69} 
it was shown that the minimum star height 
among all regular expressions denoting $L_n$ equals 
$\lfloor \log (n+1) \rfloor$,
so the star height of the second expression is optimal.
The authors suspect that this language family exhibits 
a trade-off in the sense that the  
regular expressions for~$L_n$ cannot be simultaneously
short and of low star height.  
\end{example} 

Perhaps the earliest reference
mentioning the influence of the elimination ordering 
is from 1960. 
In~\cite{McNYa60}, they proposed to identify the states 
that ``bear the most traffic,'' i.e., 
those vertices in the underlying graph with the highest degree, 
and to eliminate these states at last.
Since then, various heuristics for computing elimination orderings
that yield short regular expressions
have been proposed in the literature. 
In~\cite{LRS04}, a simple greedy heuristic was devised.  It was
proposed to assign a measure to each state, and this measure is
recomputed each time when a state is eliminated. This measure
indicates the priority in which the states are eliminated.  Observe
that eliminating a state tends to introduce new arcs in the digraph
underlying the automaton.  Thus we can order the states by a measure
that is defined as the number of ingoing arcs times the number of
outgoing arcs. In~\cite{DeMo04} a refined version of the same idea is
proposed, which takes also the lengths of the intermediate expressions
into account, instead of just counting the ingoing and outgoing arcs.
Later, a different strategy for accounting the priority 
of a state was suggested: as measure function, simply 
take the number of cycles passing through a state. 
There are some automata, where this heuristic 
outperforms the one we previously described,
but on most random \dfa{}s the performance 
is comparable. For the heuristic based on counting the 
number of cycles, recomputing the measure after the 
elimination of each state does not make a 
big difference~\cite{MNR10}. 
Another idea is to look for simple structures in 
finite automata, such as 
bridge states~\cite{HaWo07}. A bridge state typically exists 
if the language under consideration can be written as the 
concatenation of two 
nontrivial regular languages. Unfortunately, a random 
\dfa{} almost surely contains no bridge states at all, as the 
number of states grows larger~\cite{MNR10}.
These and other heuristics were compared empirically on a large set of 
random $\dfa$s as input in~\cite{GHT09,MNR10}. 
Although there are also advanced strategies for choosing an elimination 
ordering, which have provable performance guarantees, 
the greedy heuristic from~\cite{DeMo04} performs best
in most cases.  

Beyond heuristics, we can use elimination orderings
to prove nontrivial upper bounds on the conversion of \dfa{}s 
over small alphabets into regular expressions. For the case of 
binary alphabets, a bound of $O(1.742^n)$ was given 
in~\cite{GrHo08d}, which was then improved to~$O(1.682^n)$ 
in~\cite{EdFa12}. These bounds can be reached with state 
elimination by using appropriate elimination orderings. 
The latest record is $O(1.588^n)$, 
and the algorithm departs from pure 
state elimination, see~\cite{GrHo14}.

\begin{theorem}\label{thm:binary-dfa-re}
  Let $n\ge 1$ and~${A}$ be an $n$-state $\dfa$ over a binary
  alphabet.  Then size $O\left(1.588^n\right)$ is sufficient for a
  regular expression describing~$L({A})$. 
\end{theorem}

Similar bounds, but with somewhat larger constants in place
of~$1.588$, can be derived for larger alphabets. Moreover, the same
holds for \nfa s having a comparably low density of transitions. 

We sketch how to establish a simpler upper bound than this, which
after all gives $o(4^n)$ for all alphabets of constant size.  To get
things going, assume that we want to determine $L_{jk}^U$, and that
the underlying sub-graph induced by~$U$ falls apart into two mutually
disconnected sub-graphs~$A$ and~$B$.  Then on reading a word $w$, the
automaton goes from $j$ to $k$ \emph{either} through $A$ \emph{or}
through $B$, and thus~$L_{jk}^U = L_{jk}^A \cup L_{jk}^B$, and this is
reflected by the regular expressions computed using state elimination.
In particular, if the sub-graph induced by $U$ is an independent set,
i.e., a set of isolated vertices, in the underlying graph,
then~$L_{jk}^U = \bigcup_{i\in U} L_{jk}^{\{i\}}$.  In this case, the
blow-up factor incurred by eliminating~$U$ is linear in~$|U|$, instead
of exponential in $|U|$.
For a \dfa~$A$ over constant alphabet, the underlying 
graph has a linear number of edges. 
It is known that such graphs have an 
independent set of size~$cn$, where~$c$ is a constant 
depending on the number of edges. Suppose that~$U$ is 
such an independent set. Then we partition the state 
set of~$A$ into an ``easy'' part $U$ and a ``hard'' part 
$Q \setminus U$. Eliminating~$U$ increases the size of 
the intermediate expressions by a factor linear in~$|U|$. 
Thereafter, eliminating the remaining $(1-c)n$ states may 
incur a size blow-up by a factor of~$4^{(1-c)n}$. Altogether,
this gives a regular expression of alphabetic width 
in~$|\Sigma|\cdot o(4^n)$ for~$L(A)$.

Let us again take a look at an example.

\begin{example}\label{ex:independent-set} 
  For illustrating the above said, consider the language
  \[L_3=(a_1b_1)^* \shuffle (a_2b_2)^* \shuffle (a_3b_3)^*,\] where
  the \emph{interleaving}, or \emph{shuffle}, of two languages~$L_1$
  and~$L_2$ over alphabet~$\Sigma$ is
  \[L\shuffle M = \{\,w\in \Sigma^*\mid\mbox{$w\in x\shuffle y$ for
    some $x\in L$ and $y\in M$}\,\} ,\] and the interleaving
  $x\shuffle y$ of two words~$x$ and~$y$ is defined as the set of all
  words of the form $x_1y_1x_2y_2\cdots x_ny_n$, where $x=x_1x_2\cdots
  x_n$,\ $y=y_1y_2\cdots y_n$ with $x_i,y_i\in \Sigma^*$, for $n\geq
  1$ and $1\leq i\leq n$. Note that in this definition, some of the
  sub-words~$x_i$ and~$y_i$ can be empty.

  The language $L_3$ can be accepted by a partial \dfa{} over the
  state set $\{0,1\}^3$, and whose transition function is given such
  that input~$a_i$ sets the $i$th bit left of the rightmost bit of the
  current state from~$0$ to~$1$, and input~$b_i$ resets the $i$th bit,
  again counting from right to left, of the current state from~$1$
  to~$0$.  All other transitions are undefined.
  The initial state is $000$, which is also the single final state.
  Notice that the graph underlying this automaton is the
  $3$-dimensional cube, with~$8$ vertices---see
  Figure~\ref{fig:hypercube}.
\begin{figure}
  \centering
\includegraphics[scale=0.65]{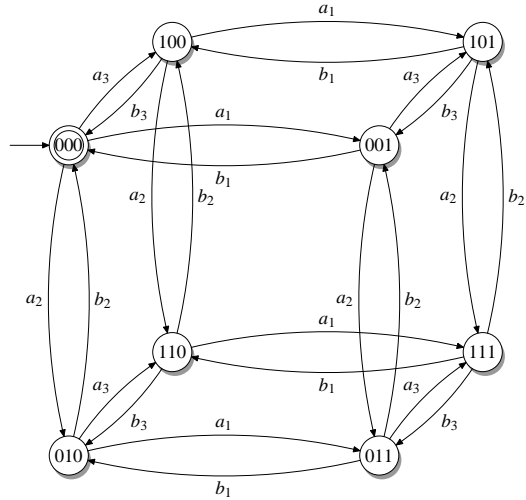}
\caption{Automaton accepting the language $L_3=(a_1b_1)^* \shuffle
  (a_2b_2)^* \shuffle (a_3b_3)^*$. The underlying graph is the
  $3$-dimensional cube.}
  \label{fig:hypercube}
\end{figure}
Generalizing this example to $d \ge 3$, the underlying graph of~$L_d$
is the $d$-dimensional hypercube, with~$2^d$ many vertices.

It is well known that the $d$-dimensional hypercube is $2$-colorable,
and thus has an independent set that contains at least half of the
vertices. Eliminating this independent set before the other vertices
yields a regular expression of alphabetic width $O(n\cdot 2^n)$, which
is way better than the trivial bound of $O(4^n)$.
\end{example} 

We present another application of this idea. Planar finite automata
are a special case of finite automata, which were first studied
in~\cite{BoCh76}.  To convert a planar finite automaton into a regular
expression, one can look for a small set of vertices, whose removal
leaves to mutually disconnected sub-graphs with vertex sets $A$ and
$B$. Then again, we have $L_{jk}^U = L_{jk}^A \cup L_{jk}^B$, and this
is reflected by the regular expressions computed by state
elimination. Since the sub-graphs induced by $A$ and $B$ are again
planar, one can apply the trick recursively. Also for this special
case, tight upper and lower bounds were found
recently~\cite{EKSW04,GrHo08,GrHo13}.

\begin{theorem}
  Let $n\ge 1$ and~${A}$ be an $n$-state planar \dfa\ or \nfa\ over
  alphabet~$\Sigma$. Then size $|\Sigma|\cdot 2^{O(\sqrt{n})}$ is sufficient for
  a regular expression describing~$L({A})$. Such an expression can be
  constructed by state elimination.
\end{theorem} 

Taking this idea again a step further, one can arrive at a parametrization 
where the conversion problem from finite automata to regular expressions is 
fixed-parameter tractable, in the sense that the problem is exponential 
in that parameter, but not in the size of the input. 
Recall that we have introduced the concept of cycle rank of a digraph 
in the course of discussing the star height in Section~\ref{sec:measures}. 
Now for a digraph $D$, let $D^{\mathrm{sym}}$ denote the symmetric 
digraph obtained by replacing each arc in $D$ with a 
pair of anti-parallel arcs. The \emph{undirected cycle rank} 
of~$D$ is defined as the cycle rank of $D^{\mathrm{sym}}$. 
If the conversion problem from finite automata to regular expressions 
is parametrized by the undirected cycle rank of the given automaton, 
one can prove the following bound~\cite{GrHo13}:

\begin{theorem}
  Let $n\ge 1$ and~${A}$ be an $n$-state \dfa\ or \nfa\ over
  alphabet~$\Sigma$, whose underlying digraph is of \emph{undirected}
  cycle rank at most~$c$, for some $c\ge 1$.  Then size $|\Sigma|\cdot
  4^c\cdot n$ is sufficient for a regular expression
  describing~$L({A})$.  Such an expression can be constructed by state
  elimination.
\end{theorem}  

Observe that fixed-parameter tractability also holds in the sense of
computational complexity, since computing the undirected cycle rank is
fixed-parameter tractable, see, e.g.,~\cite{RRSS14}. A natural
question is now whether we can find a similar parametrization in
terms of cycle rank, instead of undirected cycle rank.  Well, there
are acyclic finite automata that require regular expressions of
super-polynomial size~\cite{EhZe76,GrJo08}. Notice that these automata
have cycle rank~$0$. Hence the best we can hope for is a
parametrization that is quasi-polynomial when the cycle rank is
bounded. One can indeed obtain such an estimate~\cite{GrHo14}, but the
method is more technical, and no longer uses only state
elimination. The upper bound in terms of directed cycle rank reads as
follows:

\begin{theorem}
  Let $n\ge 1$ and~${A}$ be an $n$-state \dfa\ or \nfa\ over
  alphabet~$\Sigma$, whose underlying 
  digraph is of cycle rank at most~$c$, for some $c\ge 1$.
  Then size $|\Sigma|\cdot n^{O(c\cdot\log n)}$ is sufficient 
  for a regular expression describing~$L({A})$.
\end{theorem}

But in the general case, the exponential blow-up when moving from
finite automata to regular expressions is inherent, that is,
independent of the conversion method. Already in the 1970s the
existence of languages~$L_n$ was shown, that admit $n$-state finite
automata, but require regular expressions of alphabetic width at least
$2^{n-1}$, for all $n\geq 1$, see~\cite{EhZe76}. Their witness
language is over an alphabet of growing size, which is quadratic in
the number of states.  Their proof technique was tailored to the
witness language involved.  The question whether a comparable size
blow-up can also occur for constant alphabet size~\cite{EKSW04} was
settled only a few years ago.  The answer was provided around the same
time by two independent groups of researchers, who worked with
different proof techniques, and gave different
examples~\cite{GeNe08,GrHo08}.

How are such lower bounds established?  We shall describe a general
method, which has been used to prove lower bounds on regular
expression size in various contexts~\cite{Ge10,GrHo08,GrHo09b,HoJa11}.
In the context of lower bounds for regular expression size, a more
convenient formulation of Lemma~\ref{lem:star-height-lemma} is the
star height lemma, which reads as follows:

\begin{lemma}
  Let $L$ be a regular language. Then $\awidthc(L) \ge
  2^{\Omega(\heightc(L))}$.
\end{lemma}

That is, the minimum regular expression size of a 
regular language is at least exponential in the minimum 
required star height.
But now this looks as if we have replaced one evil with another, since
determining the star height is eminently difficult in
general~\cite{Ki05}.  But there is an important special case, in which
the star height can be determined more easily: a {\em partial}
deterministic finite automaton is called bideterministic, if it has a
single final state, and if the \nfa\ obtained by reversing all
transitions and exchanging the roles of initial and final state is
again a partial \dfa{}---notice that, by construction, this \nfa{} in
any case accepts the reversed language.  A regular language~$L$ is
\emph{bideterministic} if there exists a bideterministic finite
automaton accepting $L$.  These languages form a proper subclass of
the regular languages.  For these languages, {\em McNaughton's
  Theorem}~\cite{McN67a} states that the star height is equal to the
cycle rank of the digraph underlying the minimal partial \dfa.

\begin{example}
 Define $K_m = \{\,w\in \{a,b\}^*\mid |w|_a
 \equiv 0 \mod m\,\}$ and $L_n = \{\,w\in \{a,b\}^*\mid |w|_b \equiv
 0 \mod n\,\}$. For simplicity, assume $m\le n$.
 It is straightforward to construct 
 deterministic finite automata with~$m$ states 
 (with $n$ states, \resp) arranged in a directed cycle
 describing the languages~$K_m$ and~$L_n$, respectively.
 By applying the standard product construction on these automata, 
 we obtain a deterministic finite automaton~$A$ accepting the language
 $K_m \cap L_n$. 
\begin{figure}[htb]
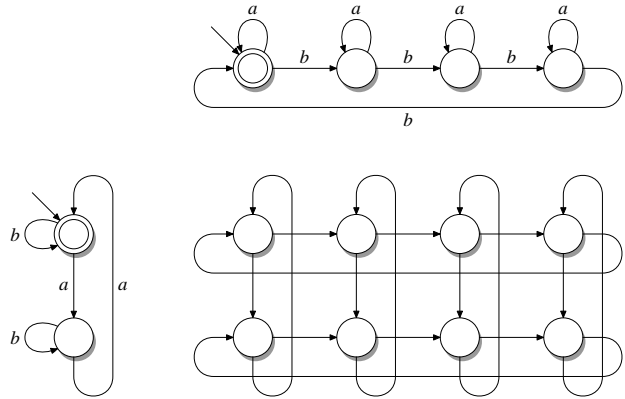

  \centering 
\begin{tabular}{ccc}
   & & \includegraphics[scale=0.65]{fig/fa2.301}\\ 
 & & \\
\includegraphics[scale=0.65]{fig/fa2.300} & &
  \includegraphics[scale=0.65]{fig/fa2.699}
\end{tabular}
\caption{Drawing of the discrete directed $(m\times n)$-torus in the
  case where $m=2$ and $n=4$, induced by the automata for the
  languages~$K_m$ and~$L_n$.}
  \label{fig:2x4-torus}
\end{figure} 
The digraph underlying automaton~$A$ is the directed discrete
torus. This digraph can be described as the Cartesian graph product of
two directed cycles, see Figure~\ref{fig:2x4-torus} for illustration.
The cycle rank of the $(m\times n)$-torus is equal to~$m$ if~$m=n$,
and equal to~$m+1$ otherwise~\cite{GrHo08}.  It is easily observed
that the automaton~$A$ is bideterministic, hence the star height
of~$L(A)$ coincides with the cycle rank of its underlying digraph.  By
invoking the star height lemma, we can derive a lower bound
of~$2^{\Omega(m)}$ on the minimum regular expression size required for
$L_m \cap K_n$.
\end{example}

For the succinctness gap between \dfa s and regular expressions over
binary alphabets, a lower bound of $2^{\Omega(\sqrt{n/\log n})}$ 
was reported in~\cite{GeNe08}, while a
parallel effort~\cite{GrHo08} resulted in an 
asymptotically tight lower
bound of~$2^{\Omega(n)}$. We have the following result:

\begin{theorem}
  Let $n\ge 1$ and~${A}$ be an $n$-state $\dfa$ or $\nfa$ over
  alphabet $\Sigma$. Then size $|\Sigma|\cdot 2^{\Theta(n)}$ is
  sufficient and necessary in the worst case for a regular expression
  describing~$L({A})$. This already holds for alphabets with
  at least two letters.
\end{theorem}
 
Recall that the notation~$2^{\Theta(n)}$ implies a lower bound
of~$c^n$, for some~$c>1$. The hidden constant in the lower bound for
binary alphabets is much smaller compared to the lower bound
of~$2^{n-1}$ previously obtained in~\cite{EhZe76} for large
alphabets. The upper bound from Theorem~\ref{thm:binary-dfa-re}
implies that~$c$ can be at most~$1.588$ for alphabets of size two.
Narrowing down the interval for the best possible~$c$ for various
alphabet sizes is a challenge for further research.

We turn our attention to interesting special cases of regular
languages, namely the \emph{finite} and the \emph{unary} regular
languages.  Here, the situation is significantly different, as we can
harness specialized techniques which are more powerful than state
elimination.  Also, finite and unary languages have star height at
most~$1$, and thus more tailored techniques than the star height lemma
are needed to establish lower bounds. Indeed, the case of finite
languages was already addressed in the very first paper on the
descriptional complexity of regular expressions~\cite{EhZe76}.  They
give a specialized conversion algorithm for finite languages, which is
different from the state elimination algorithm.  Their results imply
that every $n$-state \dfa\ accepting a finite language can be
converted into an equivalent regular expression of size~$n^{O(\log
  n)}$. The method is quite interesting, since it is not based on
state elimination, but rather on a clever application of the repeated
squaring trick. They also provide a lower bound of $n^{\Omega(\log
  \log n)}$ when using an alphabet of size $O(n^2)$.  The challenge of
tightening this gap was settled more than thirty years later
in~\cite{GrJo08}, where a lower bound technique from communication
complexity is adapted, which originated in the study of monotone
circuit complexity.

\begin{theorem}
  Let $n\ge 1$ and~${A}$ be an $n$-state \dfa\ or \nfa\ over
  alphabet~$\Sigma$ accepting a finite language.
  Then size $|\Sigma|\cdot n^{\Theta(\log n)}$ is sufficient and necessary in the
  worst case for a regular expression describing~$L({A})$. This still
  holds for constant alphabets with at least two letters.
\end{theorem}

The case of unary languages was discussed
in~\cite{Ga11,Ma02,To09}. Here the main idea is that one can exploit
the simple cycle structure of unary \dfa s and of unary \nfa s in
Chrobak normal form~\cite{Ch86b}. In the case of \nfa s, elementary
number theory helps to save a logarithmic factor of the quadratic
upper bound~\cite{Ga11}.  The main results are summarized in the
following theorem.

\begin{theorem}
  Let $n\ge 1$ and~${A}$ be an $n$-state \dfa\ over a unary
  alphabet. Then size $\Theta(n)$ is sufficient and necessary in the
  worst case for a regular expression describing~$L({A})$. When
  considering \nfa{}s, the upper bound changes to $O(n^2/\log n)$.
\end{theorem}

The tight bounds for the conversion of unary \nfa s to regular
expressions thus remain to be determined. The conversion problem has
been studied also for a few other special cases of finite
automata. Examples include finite automata whose underlying digraph is
an acyclic series-parallel digraph~\cite{MoRe09}, Thompson
digraphs~\cite{GPWZ04}, and digraphs induced by Glushkov
automata~\cite{CaZi00}.

\paragraph*{Acknowledgments.} Thanks to Sebastian Jakobi 
for helpful comments and suggestions on an earlier draft 
of this paper.


\def\lastmodified{23.12.2008}\def\id#1{#1}

\end{document}